\documentclass[pra,showpacs,preprintnumbers,amsmath,amssymb]{revtex4}
\usepackage{bm}
\usepackage{amsmath}
\usepackage[dvips]{graphicx}
\begin{document}
\bibliographystyle{apsrev}

\title{Calculation of radiative corrections to the effect of
parity nonconservation in heavy atoms}

\author{A.I.Milstein}
\email[Email:]{A.I.Milstein@inp.nsk.su} \affiliation{Budker
Institute of Nuclear Physics, 630090 Novosibirsk, Russia}
\author{O.P.Sushkov}
\email[Email:]{sushkov@phys.unsw.edu.au} \affiliation{School of
Physics, University of New South Wales, Sydney 2052, Australia}
\author{I.S.Terekhov}
\email[Email:]{I.S.Terekhov@inp.nsk.su} \affiliation{ Novosibirsk
 University, 630090 Novosibirsk, Russia}

\date{\today}
\begin{abstract}
We calculate the self-energy and the  vertex radiative corrections to the
effect of parity nonconservation in heavy atoms.
 The sum of the corrections is of the form ${\cal A}\ln(\lambda_C/r_0)+{\cal
B}$, where ${\cal A}$ and ${\cal B}$ are functions of $Z\alpha$,
and $\lambda_C$ and $r_0$ are the Compton wavelength and the nuclear
radius, respectively.
The function ${\cal A}$ is calculated exactly in $Z\alpha$ and the
function ${\cal B}$  is calculated in the leading order.
In the leading order ${\cal A}\propto \alpha(Z\alpha)^2$ and
${\cal B}\propto \alpha(Z\alpha)$.
The sum of the corrections is $-0.85\%$ for Cs and
$-1.48\%$ for Tl. Using these results we have performed analysis
of the experimental data on atomic parity nonconservation. The
values obtained for the nuclear weak charge,
$Q_W=-72.81(28)_{exp}(36)_{theor}$ for Cs and
$Q_W=-116.8(1.2)_{exp}(3.4)_{theor}$ for Tl,
agree with predictions of the standard model within $0.6\sigma$.
As an application of our approach we have also calculated
dependence of the Lamb shift on the finite nuclear size.
\end{abstract}
 \maketitle

\section{Introduction}
Atomic parity nonconservation (PNC) has now been measured in
bismuth \cite{Bi}, lead \cite{Pb}, thallium \cite{Tl}, and cesium
\cite{Cs}. Analysis of the data provides an important test of the
standard electroweak model and imposes constraints on new physics
beyond the model, see Ref. \cite{RPP}. The analysis is based on
the atomic many-body calculations for Tl, Pb, and Bi \cite{Dzuba1}
and for Cs \cite{Dzuba2,Blundell} (see also more recent Refs.
\cite{Kozlov,DFG}). Both the experimental and the theoretical
accuracy is best for Cs. Therefore,  this atom provides the most
important information on the standard model in the low-energy
sector.

In the many-body calculations  \cite{Dzuba1,Dzuba2,Blundell} the
Coulomb interaction between electrons was taken into account,
while the magnetic interaction was neglected. The contribution of
the magnetic (Breit) electron-electron interaction was calculated
in the papers \cite{Der,DHJS}. It proved to be much larger than a
naive estimate, and it  shifted the theoretical prediction for PNC
in Cs.

Radiative correction to the nuclear weak charge due to
renormalization from the scale of the W-boson mass down to zero
momentum has been calculated long time ago, see Refs.
\cite{Mar1,Mar2}. This correction is always included in the
analysis of data. However, another important class of radiative
corrections was omitted in the analysis of atomic PNC. This fact has
been pointed out in work \cite{Sushkov} that demonstrated
that there are corrections $\sim Z\alpha^2$ caused by the
collective electric field of the nucleus. Here Z is the nuclear charge and
$\alpha$ is the fine structure constant. The simplest correction
of this type is due to the Uehling potential. It has been
calculated numerically in Ref. \cite{W} and analytically in our
paper \cite{MiSu}. In that paper \cite{MiSu} we have also analyzed
the general structure of the radiative corrections caused by the
collective electric field.
It has been shown that, as well as  the usual perturbative parameter
$Z\alpha$, there is an additional parameter $\ln(\lambda_C/r_0)$
where $\lambda_C$ is the electron Compton wavelength and $r_0$ is
the nuclear radius.

In the present work we consider  radiative corrections to the
atomic PNC effect due to electron self-energy and vertex. The
total correction is of the form ${\cal A}\ln(\lambda_C/r_0)+{\cal
B}$, where ${\cal A}$ and ${\cal B}$ are functions of $Z\alpha$.
We calculate the function ${\cal A}$ exactly in $Z\alpha$ and the
function ${\cal B}$  in the leading order. In the leading order
 ${\cal A}\propto \alpha(Z\alpha)^2$ and ${\cal B}\propto
\alpha(Z\alpha)$. Results of the leading order calculations have
been reported in Ref. \cite{Mil}.
Using results of our calculations we reanalyze the experimental
data for atomic PNC. Agreement with the standard model is excellent.

As an application of our approach we have also calculated the dependence
of the Lamb shift on the finite nuclear size. Agreement of our
analytical formula with results of previous computations
\cite{CJS93,JS} is perfect. Structure of the present paper is
following. In Sec. \ref{General} we discuss the general structure
of the PNC amplitude. In Secs. \ref{Masslog} and \ref{Vertexlog} we
calculate  logarithmic parts of the self-energy  and the vertex
contributions, respectively. In
Sec.\ref{PNClinear} we calculate the linear in $Z\alpha$
correction to the PNC amplitude. In Sec. \ref{FNS} we calculation
the dependence of the Lamb shift on the finite nuclear size.
Finally in Sec. \ref{Conclusion} we analyze experimental data on
atomic PNC and present our conclusions.

\section{General structure of the PNC amplitude}\label{General}

\begin{figure}[h]
\centering
\includegraphics[height=90pt,keepaspectratio=true]{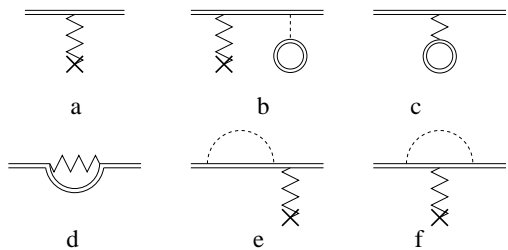}
\caption{\it (a) Leading contribution to the PNC matrix element.
(b-f) One loop radiative corrections. The double line is the exact
electron Green's function in the Coulomb field of the nucleus, the
cross denotes the nucleus, the zigzag and the dashed lines denote
Z-boson and photon, respectively.} \label{Fig1}
\end{figure}

The strong relativistic enhancement is a special property of the
atomic PNC effect. The relativistic enhancement factor is
proportional to $R \sim (\lambda_C/Z\alpha r_0)^{2(1-\gamma)}$,
where $\gamma=\sqrt{1-(Z\alpha)^2}$. The factor is $R \approx$3
for Cs and $R \approx 9$ for Tl, Pb, and Bi \cite{Kh}. For nuclear
radius we use the formula
\begin{equation}
\label{r0} r_0 \approx 1.1 \ A^{1/3}\ fm \approx 1.5 \ Z^{1/3} \
fm.
\end{equation}
The relativistic enhancement factor $R$ is divergent at $r_0 \to 0$.
This makes the relativistic behavior of PNC very much different
from the behavior of the hyperfine structure.
The logarithmic enhancement of radiative corrections mentioned
above is closely related to the existence of the factor $R$. The
Feynman diagram for the leading contribution  to the PNC matrix
element between $p_{1/2}$ and $s_{1/2}$ states as well as diagrams
for one loop radiative corrections are shown in Fig.\ref{Fig1}(a)
and Fig.\ref{Fig1}(b-f), respectively.
Strictly speaking the diagrams in Figs.\ref{Fig1}(b-f) are not quite well
defined because they describe the matrix element between states with
different energies, so it is not clear what energy corresponds to the
external legs.
 However the external legs describe states of the external atomic
electron, say $6s$, $6p$, $7s$,... states in Cs, that have very
small binding energies of the order of $m\alpha^2$ ($m$ is the
electron mass). Therefore, the uncertainty in the definition of
the diagrams appears only in the order $\alpha^3(Z\alpha)$ which
we do not consider in the present work.
\noindent The diagram Fig.\ref{Fig1}(b) corresponds to a
modification of the electron wave function because of the vacuum
polarization. This correction calculated analytically in Ref.
\cite{MiSu} reads
\begin{equation} \label{db}
\delta_{b}=\alpha\left(\frac{1}{4}Z\alpha+
\frac{2(Z\alpha)^2}{3\pi\gamma}
\left[\ln^2(b\lambda_C/r_0)+f\right]\right),
\end{equation}
where $b=\exp(1/(2\gamma)-C-5/6)$, $C\approx 0.577$ is the Euler
constant, and $f\sim 1$ is some smooth function of $Z\alpha$
independent of $r_0$. Hereafter we denote by $\delta$ the relative
value of the correction. So, Eq. (\ref{db}) represents the ratio
of diagrams Fig.\ref{Fig1}(b) and Fig.\ref{Fig1}(a).

The renormalization of the nuclear weak charge $Q_W$ from the
scale of the W-boson mass down  to $q=0$ was performed in Refs.
\cite{Mar1,Mar2}. However, as has been pointed out in Ref.
\cite{MiSu}, atomic experiments correspond to $q\sim 1/r_0\sim
30$MeV, see also Ref. \cite{Hor}. The correction due to renormalization from $q=0$ to
$q=1/r_0$ is described by diagrams Fig.\ref{Fig1}(c) and
Fig.\ref{Fig1}(d). It has the form \cite{MiSu}
\begin{equation}
\label{cd} \delta_{cd}=\frac{4\alpha Z}{3\pi Q_W}
(1-4\sin^2\theta_W)\ln(\lambda_C/r_0)\sim -0.1\% \, ,
\end{equation}
where $\theta_W $ is the Weinberg angle, $\sin^2\theta_W\approx
0.2230$, see Ref. \cite{RPP}.

Diagrams Fig.\ref{Fig1}(e) and Fig.\ref{Fig1}(f) correspond to the
contributions of the electron self-energy operator and the vertex
operator, respectively. Neither of these diagrams is  invariant
with respect to the gauge transformation of the electromagnetic
field. However, their sum  is gauge invariant. It has been
demonstrated in Ref. \cite{MiSu} that the correction $\delta_e
+\delta_f$ is of the form
\begin{equation}
\label{AB}
 \delta_{ef}=\delta_e +\delta_f={\cal A}\ln(b\lambda_C/r_0)+{\cal B} \quad ,
\end{equation}
  where ${\cal A}$ and ${\cal B}$ are functions of
$Z\alpha$, and the constant $b$ is defined after Eq.(\ref{db}). In
the leading approximation in the parameter $Z\alpha$ the functions
are, $A=(a_2/\pi) \ \alpha(Z\alpha)^2$ and $B=a_1 \
\alpha(Z\alpha)$. In the  work \cite{MiSu} we have also obtained
preliminary estimates for the coefficients $a_1$ and $a_2$. The
estimates were based on an assumption of analogy between the
polarization operator and the self-energy operator. This
assumption and hence the preliminary estimates proved to be wrong.
In the present work we calculate the coefficients $a_1$ and $a_2$
exactly. Moreover, we also calculate the function ${\cal A}$ {\it exactly}
in the parameter $Z\alpha$. The functions $\cal A$ and $\cal B$
are calculated by different methods. For calculation of $\cal A$
we use electron Green's functions in the Coulomb field at $r \ll
\lambda_C$ and the Feynman gauge. The non-logarithmic term ${\cal
B}$ we find using the effective operator method and the
Fried-Yennie gauge \cite{FY}. The method is based on calculation
of the low-energy scattering amplitude in the potential induced by
$Z$-boson exchange.

Let us consider the PNC matrix element between $p_{1/2}$ and
$s_{1/2}$ states. The wave function of the external electron is of
the form
\begin{equation}
\label{Dirac} \psi({\bm r})= \begin{pmatrix}
F(r)\Omega\\
iG(r)\tilde{\Omega}
\end{pmatrix}
\quad ,
\end{equation}
where $\Omega$ and $\tilde{\Omega}=-({\bm \sigma}\cdot{\bm
n})\Omega$ are spherical spinors\cite{BLP}. Calculation of the
weak interaction matrix element in the leading approximation gives
\cite{Kh}
\begin{equation}
\label{pnc}
<p_{1/2}|(2\sqrt{2})^{-1}G_FQ_W\rho(r)\gamma_5|s_{1/2}>_0=M_0\propto
(F_sG_p-G_sF_p)|_{r=r_0}.
\end{equation}
Here $G_F$ is the Fermi constant, and $\rho(r)$ is the density of
the nucleus normalized as $\int\! \rho(r)d{\bm r} =1$. At small
distances, $r \ll Z\alpha\lambda_C$, the electron mass can be
neglected compared to the Coulomb potential energy, and solution
of the Dirac equation for the radial wave functions reads
\begin{equation}
\label{fg1} F= Nr^{\gamma-1},\ \ \
G=N\frac{Z\alpha}{\kappa-\gamma}r^{\gamma-1},
\end{equation}
where $\kappa=-1$ for $s_{1/2}$-state and $\kappa=1$ for
$p_{1/2}$-state,  and $N$ is some constant dependent on the
wave-function behavior at large distances ($r\sim a_B$) \cite{Kh}.
Calculation of the constant $N$ is a very complex many-body
problem, see Refs.\cite{Dzuba2,Blundell}. Fortunately, the
constant is cancelled out from the relative correction which we
consider. Due to Eq. (\ref{fg1}) the matrix element (\ref{pnc}) is divergent
at $r_0 \to 0$ , $M_0\propto r_0^{2\gamma-2}$. This leads to the
strong relativistic enhancement factor mentioned above.

 The correction to $M_0$ related to the diagram Fig.\ref{Fig1}(e) can be
obtained as follows.  The electron self-energy operator
${\bm\Sigma}$ being substituted to the Dirac equation,
$m \to m+{\bm\Sigma}$, leads to the Lamb shift of
energy levels and to the modification of electron wave
functions, see, e.g., Ref. \cite{BLP}. The wave function
modification influences the matrix element of the weak
interaction. We will call the Dirac equation with account of the
self-energy operator the modified Dirac equation.

It is convenient to search for solution of the modified Dirac
equation in the form (\ref{Dirac}) with substitutions $F\to
F(1+F^{(1)})$, $G\to G(1+G^{(1)})$, where $F^{(1)}$ and $G^{(1)}$
are corrections due to the self energy. Using this form in
Eq.(\ref{pnc}) together with (\ref{fg1}) we obtain the following
relative correction
 to the PNC matrix element
\begin{eqnarray}\label{delta}
\delta = M^{(1)}/M^{(0)}
=\left\{\frac{1+\gamma}{2}\,\left[F_s^{(1)}+{G}_p\!\!\!{}^{(1)*}\right]+
\frac{1-\gamma}{2}\,\left[G_s^{(1)}+{F}_p\!\!\!{}^{(1)*}\right]
\right\}\Big|_{r=r_0}\quad .
 \end{eqnarray}
Similar to Eqs.(\ref{Dirac}),(\ref{fg1}),   we represent the
self-energy term in the Dirac equation $\Psi^{\Sigma}={\bm\Sigma}\psi$ as
\begin{equation}\label{SWF}
\Psi^{\Sigma}=
\begin{pmatrix}
  {F}^{(2)}\Omega \\
 i{G}^{(2)}\dfrac{Z\alpha}{\kappa-\gamma}\tilde{\Omega}
\end{pmatrix}\,r^{\gamma-2}\quad .
\end{equation}
In the next Section we will show that $F^{(2)}$ and $G^{(2)}$ are
independent of $r$ at $r\ll Z\alpha \lambda_C$. The functions
$F^{(1)}$ and $G^{(1)}$ satisfy the following equations
\begin{eqnarray}\label{equations}
&&   \frac{1}{\gamma+\kappa}\frac{\partial F^{(1)}}{\partial
x}-F^{(1)}+G^{(1)}=-\frac{1}{Z\alpha}G^{(2)}\nonumber \\
&&   \frac{1}{\kappa-\gamma}\frac{\partial G^{(1)}}{\partial
x}-F^{(1)}+G^{(1)}=-\frac{1}{Z\alpha}F^{(2)}\quad ,
\end{eqnarray}
where $x=-\ln(r/\lambda_C)$. The solution of Eqs. (\ref{equations}) reads
\begin{eqnarray}\label{solution}
&& F^{(1)}= \frac{Z\alpha x}{2\gamma}\,[G^{(2)}-F^{(2)}]+
\frac{1}{4Z\alpha\gamma}\,[(\kappa+\gamma)G^{(2)}-(\kappa-\gamma)F^{(2)}]+a
\, ,\nonumber\\
&& {G}^{(1)}= \frac{Z\alpha x}{2\gamma}\,[G^{(2)}-F^{(2)}]-
\frac{1}{4Z\alpha\gamma}\,[(\kappa+\gamma)G^{(2)}-(\kappa-\gamma)F^{(2)}]+a
\, ,
\end{eqnarray}
where $a$ is some constant. Note that in the leading logarithmic approximation
(the terms proportional to $x$) $F^{(1)}=G^{(1)}$. Substituting
(\ref{solution}) into Eq.(\ref{delta}), we find
\begin{eqnarray}\label{deltasigma}
\delta_e&=& \frac{Z\alpha
L}{2\gamma}\left[G_s^{(2)}+{G}_p\!\!\!{}^{(2)*}-F_s^{(2)}-
{F}_p\!\!\!{}^{(2)*}\right]+{const.} \,
 \end{eqnarray}
Here $L=\ln(b\lambda_C/r_0)$, and by $const$ we denote
$Z\alpha$-dependent terms that do not contain the large logarithm $L$.
It will be shown below that $F_p^{(2)}=-G_s^{(2)}$ and
$G_p^{(2)}=-F_s^{(2)}$. Hence, the correction $\delta_e$ is of the
form
\begin{eqnarray}\label{deltasigma0}
\delta_{e}= \frac{Z\alpha
}{\gamma}\mbox{Re}\left[G_s^{(2)}-F_s^{(2)}\right]L+ {const}\, .
 \end{eqnarray}
We will demonstrate that $G_s^{(2)}$ and $F_s^{(2)}$ are  odd
functions of $Z\alpha$. Therefore, the factor before the logarithm
$L$ is  an even function of $Z\alpha$ and hence the leading term
of $Z\alpha$-expansion of this function is
proportional to $\alpha(Z\alpha)^2$. The $const$ in
Eq.(\ref{deltasigma0}) can be determined from the condition that
the correction to the wave function is orthogonal to the
solution of the non-modified Dirac equation. Therefore, the main
contribution to the $const$ comes from the distances $r\sim
\lambda_C$, and hence the leading term  of $Z\alpha$-expansion of the
$const$ is proportional to $\alpha(Z\alpha)$. The correction $\delta_f$
corresponding to the diagram  in Fig.\ref{Fig1}(f) has the same properties as
$\delta_e$.

\section{Logarithmic contribution of the self-energy operator}\label{Masslog}

At $r \ll \lambda_C$ one can neglect the electron mass in the
propagator and write down $\Psi^{\Sigma}={\bm\Sigma}\psi$ as
\begin{eqnarray}\label{PsiS}
\Psi^{\Sigma}(\bm
r_2)=2\alpha\int\limits_{-\infty}^{\infty}\!\!d\epsilon\int d\bm
r_1\gamma_\mu {\cal G}(\bm r_2,\bm r_1|\,i\epsilon) \gamma^\mu\,
D(\bm r_2,\bm r_1|i\epsilon)\psi(\bm r_1) \quad ,
\end{eqnarray}
where ${\cal G}$ is the electron Green's function in the Coulomb
field, $g_{\mu\nu}D$ is the photon Green's function in the Feynman
gauge. In (\ref{PsiS}) we have deformed the contour of integration
over the energy $\epsilon$ in such a way that it coincides with
the imaginary axis. According to the integral representation
derived in \cite{MS82}, the Green's function is of the form
\begin{eqnarray}\label{Green}
&& {\cal G}(\bm r_2,\bm r_1|\,i\epsilon)= -\frac{i}{4\pi
r_1r_2}\sum_{l=1}^{\infty}\int\limits_0^{\infty} ds\exp\left[\,
2iZ\alpha\lambda s-k(r_1+r_2)\coth s\,\right] T\, , \nonumber\\
&& T=\gamma^{0}[1-(\bm \gamma\bm n_2 )(\bm \gamma\bm n_1
)]\left[\lambda\frac{y}{2}I_{2\nu}^{\prime}(y)-iZ\alpha\coth
s\,I_{2\nu}(y)\right]B +\gamma^{0}[1+(\bm \gamma\bm n_2)(\bm
\gamma\bm n_1 )]\lambda I_{2\nu}(y)A
\nonumber\\
&&-\left[\frac{k(r_2-r_1)}{2\sinh^2s}(\bm \gamma , \bm n_1 +\bm
n_2 )B+\coth s (\bm \gamma , \bm n_2 -\bm
n_1)A\right]I_{2\nu}(y) \, , \nonumber\\
&&y=\frac{2k\sqrt{r_1r_2}}{\sinh s}\, ,\quad
A=l\frac{d}{dx}(P_l(x)+P_{l-1}(x))\quad ,\quad
B=\frac{d}{dx}(P_l(x)-P_{l-1}(x))\quad .
\end{eqnarray}

Here $k=|\epsilon|$ , $\lambda=\epsilon/k$ , $\bm n_{1,2}=\bm
r_{1,2}/r_{1,2}$, $x={\bm n}_1\cdot{\bm n}_2$ , $P_l(x)$ is the
Legendre polynomial, $I_{2\nu}$ is the modified Bessel function of
the first kind, $\nu=\sqrt{l^2-(Z\alpha)^2}$. Using relation
$$\int (\bm n_1\times \bm n_2)\Phi(\bm n_1 \cdot\bm n_2)
 d\bm n_1=0\quad ,$$
that is valid for arbitrary function $\Phi(x)$, and
Eqs.(\ref{Dirac}), (\ref{fg1}), (\ref{Green}), and(\ref{PsiS}) we
get for $s$-wave
\begin{eqnarray}\label{FG2}
&&\begin{pmatrix} F_s^{(2)}\\ G_s^{(2)} \end{pmatrix}
 =
\frac{i\alpha
r_2^{1-\gamma}}{\pi}\sum_{l=1}^{\infty}\sum_{\lambda=\pm
1}\int\limits_0^{\infty}\!\!d\epsilon \int d\bm r_1\,
r_1^{\gamma-2} \int\limits_0^{\infty}\!\! ds  q\exp\left[\,
2iZ\alpha\lambda s-k(r_1+r_2)\coth s\,\right]D(\bm r_2,\bm
r_1|i\epsilon)
\begin{pmatrix}
f_s\\
g_s
\end{pmatrix}
\, ,\nonumber \\
&&
f_s=(1+x)B\left[\lambda\frac{y}{2}I_{2\nu}^{\prime}(y)-iZ\alpha\coth
s\,I_{2\nu}(y)\right] +(1-x)A\lambda I_{2\nu}(y)
\nonumber\\
&&-\frac{iZ\alpha}{1+\gamma}\left[(1+x)B\frac{k(r_2-r_1)}{2\sinh^2s}-(1-x)A\coth
s\right]I_{2\nu}(y) \, , \nonumber\\
&&
g_s=(1+x)(1-2x)B\left[\lambda\frac{y}{2}I_{2\nu}^{\prime}(y)-iZ\alpha\coth
s\,I_{2\nu}(y)\right] -(1-x)(1+2x)A\lambda I_{2\nu}(y)
\nonumber\\
&&-\frac{iZ\alpha}{1-\gamma}\left[(1+x)B\frac{k(r_2-r_1)}{2\sinh^2s}+(1-x)A\coth
s \right]I_{2\nu}(y) \, .
\end{eqnarray}
Performing similar calculations for p-wave we find that
$F_p^{(2)}=-G_s^{(2)}$ and $G_p^{(2)}=-F_s^{(2)}$. The explicit
form of the photon Green's function $D$ reads:
\begin{equation}\label{D}
D(\bm r_2,\bm r_1|i\epsilon)=-\frac{e^{-kR}}{4\pi R}\quad ,\quad
R=|\bm r_2-\bm r_1| \quad .
\end{equation}
Using Eqs.(\ref{FG2}), (\ref{D}),(\ref{deltasigma0}) as well as the
substitution  $ \bm r_1=r_2 \bm\rho$ , $\epsilon =E/r_2$ we obtain
the following expression for the self-energy contribution to the
function ${\cal A}$ (see, Eq.(\ref{AB}))
\begin{eqnarray}\label{deltasigma1}
{\cal A}_{SE}&=& -\frac{\alpha}{\pi^2\gamma}\sum_{l=1}^{\infty}
\int\limits_0^{\infty}\!\!dE \int \frac{d\bm \rho\,
\rho^{\gamma-2}}{R}\, \int\limits_0^{\infty}\!\! ds
\exp\left[-E(\rho+1)\coth s-E R\right]\,\,
(\Delta_1+\Delta_2)\quad ,\nonumber \\
\Delta_1
&=&x(1+x)B\left[\frac{y}{2}I_{2\nu}^{\prime}(y)(Z\alpha)\sin\phi-
(Z\alpha)^2\cos\phi\coth s\,I_{2\nu}(y)\right]
+(Z\alpha)(1-x^2)A\sin\phi\,I_{2\nu}(y) \, ,\nonumber\\
\Delta_2&=& \cos\phi\left[(1+x)B
\frac{E\gamma(1-\rho)}{2\sinh^2s}+(1-x)A\coth s\,
\right]I_{2\nu}(y) \, .
\end{eqnarray}
Here $\phi=2Z\alpha s$ , $R=\sqrt{\rho^2-2\rho x +1}$,
$y=2E\sqrt{\rho}/\sinh s$. It is obvious from this formula that
${\cal A}_{SE}$ is an even function of $Z\alpha$.

\begin{figure}[h]
\centering \vspace{10pt}
\includegraphics[height=35pt,keepaspectratio=true]{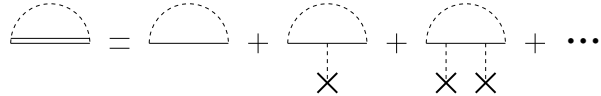} \caption{\it
{The electron self energy expanded in powers of the Coulomb field.
The solid line is the free electron Green's function, the cross
denotes the nucleus, and the dashed line denotes the photon. }}
\label{Fig2}
\end{figure}

Equation (\ref{deltasigma1}) is not quite correct. The point is that the
integrand in this equation has been derived for $\bm r_1\ne \bm
r_2$. However, there are also terms containing $\delta$-function
$\delta(\bm r_2-\bm r_1)$ and derivative of this
$\delta$-function. To account for these terms we  represent the
self-energy operator as a series in powers of the Coulomb field of
the nucleus, $\bm\Sigma=\bm\Sigma_0+\bm\Sigma_1+\bm\Sigma_2+...$,
see Fig.\ref{Fig2}.
\noindent
 The first two terms in the series, $\bm\Sigma_0$ and
$\bm\Sigma_1$, require the ultraviolet  regularization.
It means that they contain terms with the $\delta$-function
and derivative of the $\delta$-function.
Both $\bm\Sigma_0$ and $\bm\Sigma_1$ depend on the parameter of the
ultraviolet regularization. However,  the contribution of
$\bm\Sigma_{01}=\bm\Sigma_0+\bm\Sigma_1$ to ${\cal A}_{SE}$ is
independent of the regularization parameter because of the Ward
identity. The operators $\bm\Sigma_n$ ($n\geq 2$) do not require
any regularization. Note that $\Delta_1$ in Eq.(\ref{deltasigma1})
corresponds to  the contribution of the operators $\bm\Sigma_{2n+1}$,
while $\Delta_2$ corresponds to that of the operators
$\bm\Sigma_{2n}$. Let us represent the self energy operator as
$\bm\Sigma=\bm\Sigma_{01}+ \bm\Sigma^{(n>1)}$. Corresponding
contributions to the function ${\cal A}_{SE}$ are ${\cal
A}_{SE}^{(01)}$ and  ${\cal A}_{SE}^{(n>1)}$, hence
\begin{equation}
\label{ase} {\cal A}_{SE}= {\cal A}_{SE}^{(01)}+ {\cal
A}_{SE}^{(n>1)}.
\end{equation}
 The terms singular at $\bm r_1= \bm r_2$
contribute only to ${\cal A}_{SE}^{(01)}$. To overcome the problem
with singularity we calculate ${\cal A}_{SE}^{(01)}$ using
momentum representation and the standard covariant regularization of
$\bm\Sigma_0$ and $\bm\Sigma_1$ .
A simple calculation  in the momentum space gives the following expression
for $\bm\Sigma_0$
\begin{eqnarray}\label{S0}
\bm\Sigma_0(\bm p_2,\,\bm p_1)=
-\frac{\alpha}{2\pi}\left[\frac{3}{4}+\frac{1}{2}\ln\left(\frac{\Lambda^2}{\bm
p_2^2}\right) \vartheta(\Lambda^2-\bm p_2^2)\right]{\hat p}_2
(2\pi)^3\delta(\bm p_2-\bm p_1) \quad ,
 \end{eqnarray}
where $\Lambda$ is the parameter of the regularization, and
$\vartheta(x)$ is the step-function. Because of the Dirac
equation, ${\hat p}\psi = (-Z\alpha/r)\gamma_0\psi$, the operator
$\bm\Sigma_0$ in the modified Dirac equation can be replaced by
\begin{eqnarray}\label{S0p}
\widetilde{\bm\Sigma}_0(\bm p_2,\,\bm p_1)= \frac{2\alpha
(Z\alpha)}{(\bm p_2-\bm
p_1)^2}\left[\frac{3}{4}+\frac{1}{2}\ln\left(\frac{\Lambda^2}{\bm
p_2^2}\right) \vartheta(\Lambda^2-\bm p_2^2)\right]\gamma^0
 \quad ,
 \end{eqnarray}
The operator (\ref{S0p}) has the same $\gamma$-matrix structure as
that of $\bm\Sigma_1$. As a result the sum of $\bm\Sigma_1$ and
$\widetilde{\bm\Sigma}_0$ is independent of $\Lambda$ :
\begin{eqnarray}\label{S01p}
{\bm\Sigma}_{01}(\bm p_2,\,\bm p_1)&=& \frac{\alpha(Z\alpha)}{(\bm
p_2-\bm
p_1)^2}\left\{1+2\int\limits_0^1\!\!\int\limits_0^1\frac{dxdy}{T}
\left[x^2y^2(\bm p_2-\bm p_1)^2 +(1-x)(x\bm p_1^2+\hat p_1\hat
p_2)\right]\right\}\gamma^0
 \quad ,\nonumber\\
T&=&y(1-xy)\bm p_2^2+(1-y)(1-x +xy)\bm p_1^2-2xy(1-y)\bm
p_1\cdot\bm p_2\quad .
\end{eqnarray}
The Fourier transform of Eqs. (\ref{Dirac}),(\ref{fg1}) gives the
following result for the $s_{1/2}$ wave function
in momentum representation
\begin{equation}
\label{Diracp} \psi({\bm p})=
N\frac{4\pi\Gamma(\gamma+1)}{p^{\gamma+2}}\begin{pmatrix}
\cos(\pi\gamma/2)\Omega\\
(Z\alpha/\gamma)\sin(\pi\gamma/2)(\bm\sigma\cdot \bm n_p) \Omega
\end{pmatrix}
\quad ,
\end{equation}
where $\bm n_p = \bm p/p$. Using Eqs. (\ref{S01p}),
(\ref{Diracp}), and definition of $F^{(2)}$ and $G^{(2)}$ given in
Eq. (\ref{SWF}), we find
\begin{eqnarray}
\label{f2g2}
F_s^{(2)}&=&\gamma\frac{\cos (\pi\gamma/2)} {\sin(\pi\gamma/2)}
\int\frac{p_2^{\gamma+1}}{p_1^{\gamma+2}}
\Sigma_{01}(\bm p_2,\bm p_1)
\frac{d \bm p_1}{(2\pi)^3},\nonumber\\
G_s^{(2)}&=&\frac{(\gamma^2-1)\sin (\pi\gamma/2)}
{\gamma\cos(\pi\gamma/2)}
\int\frac{p_2^{\gamma}}{p_1^{\gamma+3}}({\bf p_1\cdot
p_2})\Sigma_{01}(\bm p_2,\bm p_1) \frac{d \bm p_1}{(2\pi)^3} \ \ .
\end{eqnarray}
Evaluation of these integrals and substitution of $F_s^{(2)}$ and
 $G_s^{(2)}$ in Eq. (\ref{deltasigma0}) (see also Eq. (\ref{AB})) gives
\begin{eqnarray}\label{A01}
{\cal A}_{SE}^{(01)}&=&-\frac{\alpha(Z\alpha)^2}{2\pi\gamma}
\left[\psi(2+\gamma)+\psi(2-\gamma)
-\psi(\frac{1+\gamma}{2})-\psi(1-\frac{\gamma}{2})
-\frac{1-\gamma}{2(1+\gamma)}\right.\nonumber\\
&&+\left.\psi(1/2)-\psi(1)-1+\frac{3}{\gamma^2}+\frac{6}{1-\gamma^2}
-\frac{3 \pi}{\gamma \sin(\pi\gamma)}\right]\quad .
\end{eqnarray}
Here $\psi(x)=d \ln\Gamma(x)/dx$. The first term of
$Z\alpha$-expansion of the function ${\cal A}_{SE}^{(01)}$ is
$-\alpha(Z\alpha)^2/\pi$. Note that even at $Z=90$ the exact value
of ${\cal A}_{SE}^{(01)}$ differs from that given by the first
term of the expansion by $10\%$ only.

\begin{figure}[h]
\centering
\includegraphics[height=180pt,keepaspectratio=true]{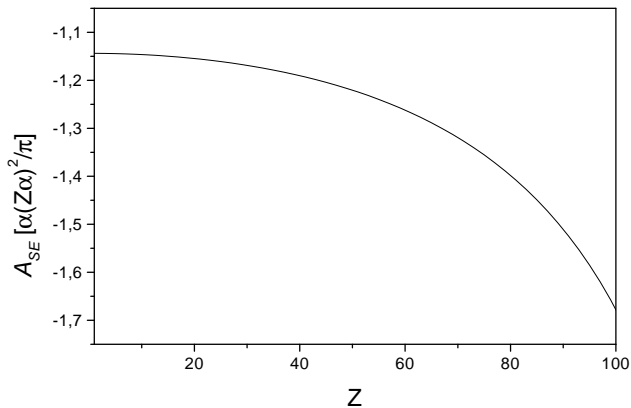}
\caption{\it The function ${\cal A}_{SE}$ calculated in all orders in
$Z\alpha$. Value of the function is given in units
$\frac{\alpha(Z\alpha)^2}{\pi}$.
} \label{Fig3}
\end{figure}

To find ${\cal A}_{SE}^{(n>1)}$ one has to subtract from
(\ref{deltasigma1}) the contribution (\ref{A01}). To do this we
subtract from the integrand in (\ref{deltasigma1}) the first term
of its $Z\alpha$-expansion at fixed $\gamma$. The point is that
$\gamma$ in the integrand comes from the wave function, but not
from the self-energy operator. Therefore, doing this subtraction
one has to consider $\gamma$ as an independent parameter. After
the subtraction we make substitutions $E=\varepsilon/2\sqrt{\rho}$
and  $\rho=\exp(2\tau)$, and take the integral over the azimuth
angle of the vector $\bm \rho$. Altogether this gives
\begin{eqnarray}\label{deltasigma2}
{\cal A}_{SE}^{(n>1)}&=&
-\frac{2\alpha}{\pi\gamma}\sum_{l=1}^{\infty}\int\limits_{-1}^{1}dx
\int\limits_0^{\infty}\!\!\int\limits_0^{\infty}\!\!\int\limits_0^{\infty}
\frac{d\varepsilon\,d\tau\,ds}{{\cal D}}\,
 \exp\left[-\varepsilon({\cal D}+\cosh\tau\coth s)\right]\,\,
\Delta\quad ,\nonumber \\
\Delta
&=&x(1+x)B\cosh(2\gamma\tau)\left\{(Z\alpha)\frac{y}{2}
\left[\sin\phi\,I_{2\nu}^{\prime}(y)
-\phi I_{2l}^{\prime}(y)\right] - (Z\alpha)^2\coth s
\left[\cos\phi\,I_{2\nu}(y)-I_{2l}(y)\right]\right\}\nonumber\\
&&+(Z\alpha)(1-x^2)A\cosh(2\gamma\tau)
\left[\sin\phi\,I_{2\nu}(y)-\phi\,I_{2l}(y)\right]\nonumber\\
&&+\left[\cos\phi\,I_{2\nu}(y)-I_{2l}(y)\right] \left[-(1+x)
B\frac{\varepsilon\gamma\sinh\tau\sinh(2\gamma\tau)}{2\sinh^2s}+
(1-x)A\coth s\,\cosh(2\gamma\tau)\right] \, .
\end{eqnarray}
Here ${\cal D}=\sqrt{\sinh^2\tau + (1-x)/2}$, and
 $y=\varepsilon/\sinh s$. The leading term of $Z\alpha$-expansion
 of ${\cal A}_{SE}^{(n>1)}$ is
 $$-\alpha(Z\alpha)^2(\pi^2-9)/(6\pi).$$
It corresponds to the contribution of the operator
$\bm\Sigma_{2}$. According to Eq. (\ref{ase}), the leading term of
$Z\alpha$-expansion of ${\cal A}_{SE}$  reads
\begin{eqnarray}\label{munu}
A_{SE}&=&-\frac{\alpha(Z\alpha)^2}{6\pi}(\pi^2-3)\quad.
\end{eqnarray}
Numerical calculation of integrals in Eq. (\ref{deltasigma2}) together with
Eq. (\ref{A01}) gives the function ${\cal A}_{SE}$ exactly in $Z\alpha$.
This function is plotted in Fig. \ref{Fig3}

\section{Logarithmic contribution of the vertex}\label{Vertexlog}

Let us consider now the logarithmic contribution of the vertex
operator (diagram (f) in Fig.\ref{Fig1}). According to the rules
of the diagram technique, the expression for the vertex correction
to the matrix element is of the form
\begin{eqnarray}\label{v1}
&&M_f=2\alpha\int\limits_{-\infty}^{\infty}\!\!\!{d\epsilon}\int\!\!\!\int\!\!\!\int\!\!\!
d\bm r_1d\bm r_2d\bm r_3\,\bar\psi_p({\bm r_2})\gamma_\mu {\cal
G}(\bm
r_2,\bm r_3|\,i\epsilon)\nonumber\\
&&\times \gamma^0\gamma^5\rho(\bm r_3){\cal G}(\bm r_3,\bm
r_1|\,i\epsilon)\gamma^\mu\psi_s(\bm r_1)D(\bm r_2,\bm
r_1|i\epsilon) \,\, ,
\end{eqnarray}
where, as above, we neglect in (\ref{v1})  the electron mass. We
use expansion of the electron Green's function in spherical waves,
see Eq. (\ref{Green}). Because of the contact nature of the weak
interaction, only the term with angular momentum $j=1/2$ is
important in the expansion, this term corresponds to $l=1$ in Eq.
(\ref{Green}). Note that $j=1/2$ dominates only in the calculation
with logarithmic accuracy. For a calculation of the constant near
the logarithm  one has to take into account all the partial waves
in spite of the contact nature of the weak interaction. Now we
calculate only the logarithmic contribution. The logarithmic
contribution to the amplitude $M_f$ comes from the region of
integration $r_1\sim r_2$, $\lambda_{C}\,\gg r_{1,2}\, \gg\,
r_3\sim r_0$, $\epsilon\,\sim\, 1/r_{1,2}$. In this case the
parameter $s$ in the integral representation Eq.(\ref{Green})  is
of the order of unity, $s\sim 1\,$, and hence the argument of the
Bessel function is small, $y\sim (r_3/r_{1,2})^{1/2} \ll 1$.
Expanding Bessel functions we transform Eq.(\ref{Green}) to the
following form
\begin{eqnarray}\label{Gc1}
&&{\cal G}(\bm r_2,\bm r_1|\, i\epsilon) = -\frac{i}{4\pi
r_1r_2\Gamma(2\gamma+1)}\int_{0}^{\infty}ds\, \exp\left[\,
2iZ\alpha\lambda s\, -\, k(r_2+r_1)\coth s\,\right]
\nonumber\\
&& \times\left(\frac{k\sqrt{r_2r_1}}{\sinh
s}\right)^{2\gamma}\Biggl\{[1-(\bm \gamma\bm n_2 )(\bm \gamma\bm
n_1 )] \gamma^{0}(\gamma\lambda-iZ\alpha\coth s) +
[1+(\bm \gamma\bm n_2 )(\bm \gamma\bm n_1 )]\gamma^{0}\lambda \nonumber \\
&& -\frac{k(r_2-r_1)}{2\sinh^2s}(\bm \gamma , \bm n_2 +\bm n_1 )
\, -\, (\bm \gamma , \bm n_2 -\bm n_1 )\coth s\, \Biggr\} \,.
\end{eqnarray}
 It is convenient to perform  integration by parts over $s$
 in the term proportional to $(r_2-r_1)\,$. After that we substitute
(\ref{Gc1}) in equation  (\ref{v1}) and take the integral over $\bm r_3$.
As a result we get
\begin{eqnarray}\label{v2}
&&M_f=\frac{\alpha\,
r_0^{2(\gamma-1)}}{2\pi^2\Gamma^2(2\gamma+1)}\int\limits_0^{\infty}\!\!d\epsilon
\int\!\!\!\int \frac{d\bm r_1d\bm r_2}{r_1r_2} \sum_{\lambda=\pm
1}\int\limits_{0}^{\infty}\!\!\!\int\limits_{0}^{\infty}\!\!ds_1ds_2\,
\exp\left[\, 2iZ\alpha\lambda (s_1+s_2)-\epsilon(r_1\coth
s_1+r_2\coth s_2)\,\right]\nonumber\\
&& \times\left(\frac{\epsilon^2\sqrt{r_1r_2}}{\sinh s_1\sinh
s_2}\right)^{2\gamma}\,\bar\psi_p({\bm
r_2})\Biggl\{\gamma^{0}[1-(\bm \gamma\bm n_1 )(\bm \gamma\bm n_2
)]\left(1-\coth s_1\coth s_2\right) +\gamma^{0}[1+(\bm \gamma\bm
n_1)(\bm \gamma\bm n_2 )]
\nonumber\\
&&\times \left[ \gamma\left(1+\coth s_1\coth
s_2\right)-iZ\alpha\lambda \left(\coth s_1+\coth
s_2\right)\right]+\lambda(\bm \gamma , \bm n_1 +\bm n_2 )
\left(\coth s_1-\coth s_2\right)\nonumber\\
&&+(\bm \gamma , \bm n_1 -\bm n_2 )\left[\gamma\lambda\left(\coth
s_1+\coth s_2\right)-iZ\alpha \left(\coth s_1\coth
s_2+1\right)\right] \Biggr\}\gamma^5\psi_s(\bm r_1)\,D(\bm r_2,\bm
r_1|i\epsilon) \, .
\end{eqnarray}
Then we substitute the photon propagator (\ref{D}) and the
electron wave functions (\ref{Dirac}), (\ref{fg1}), and find the
following expression  for the relative correction $\delta_f$
\begin{eqnarray}\label{v3}
\delta_f&=&-\frac{\alpha}{8\pi^3\Gamma^2(2\gamma+1)}
\int\limits_0^{\infty}\!\! d\epsilon \int\!\!\!\int \frac{d\bm
r_1d\bm r_2}{r_1^2r_2^2}\frac{\exp(-\epsilon R)}{R}
\sum_{\lambda=\pm
1}\int_{0}^{\infty}\!\!\!\int_{0}^{\infty}ds_1ds_2\, \exp\left[\,
2iZ\alpha\lambda (s_1+s_2)\right.\nonumber\\
&&\left.-\epsilon(r_1\coth s_1+r_2\coth s_2)\,\right]\,
\left(\frac{\epsilon^2{r_1r_2}}{\sinh s_1\sinh
s_2}\right)^{2\gamma}\,{\cal F}\, , \nonumber\\
{\cal F}&=&(1-x^2)(1-\gamma)\left[(1+\gamma)\coth s_1\coth
s_2-(1-\gamma) - iZ\alpha\lambda \left(\coth s_1+\coth
s_2\right)\right]+ 2(1-x\coth
s_1\coth s_2) \, ,\nonumber\\
R&=&\sqrt{r_1^2+r_2^2-2xr_1r_2}\quad ,\quad x=\bm n_1 \cdot\bm
n_2\quad .
\end{eqnarray}
At the next step we introduce new variables $\rho$, $t$, $E$ via
the relations $ r_1=\rho t$, $r_2=\rho/t$,
$\epsilon\sqrt{r_1r_2}=E$, perform integration over $\rho$
 with logarithmic accuracy in limits $r_0 < \rho <\lambda_C$,
 and take integrals over all the angles except of the angle between
vectors $\bm r_{1}$ and $\bm r_{2}$. This gives the following
expression for the contribution  of the vertex operator to the
function ${\cal A}$ defined in Eq.(\ref{AB}):
\begin{eqnarray}\label{v4}
{\cal A}_V&=&-\frac{2\alpha}{\pi\Gamma^2(2\gamma+1)}
\int\limits_0^\infty\!\!{dE}
\int\limits_0^\infty\!\!\frac{dt}{t}\int\limits_{-1}^1dx\,\frac{\exp(-E
R)}{R} \sum_{\lambda=\pm
1}\int\limits_{0}^{\infty}\!\!\!\int\limits_{0}^{\infty}ds_1ds_2\,
\exp\left[\,
2iZ\alpha\lambda (s_1+s_2)\right.-\nonumber\\
&&\left.-E(t\coth s_1+(1/t)\coth s_2)\,\right]\,
\left(\frac{E^2}{\sinh s_1\sinh s_2}\right)^{2\gamma}\,{\cal F}\,
.
\end{eqnarray}
Here $R=\sqrt{t^2+1/t^2-2x}$ and ${\cal F}$ is defined in
(\ref{v3}). A simple integration over $E$ leads to
\begin{eqnarray}\label{v5}
{\cal A}_V&=&-\frac{2\alpha\Gamma(4\gamma+1)}
{\pi\Gamma^2(2\gamma+1)}
\int_0^\infty\frac{dt}{t}\int_{-1}^1\,\frac{dx}{R}
\sum_{\lambda=\pm
1}\int_{0}^{\infty}\!\!\!\int_{0}^{\infty}\frac{ds_1ds_2\, \exp[\,
2iZ\alpha\lambda (s_1+s_2)]}{(\sinh s_1\sinh
s_2)^{2\gamma}}\nonumber\\
&&\times\frac{{\cal F}}{[t\coth s_1+(1/t)\coth
s_2+R]^{4\gamma+1}}\, .
\end{eqnarray}
The last integration which can be performed analytically is the
integration over $x$. The result is  rather  cumbersome:
\begin{eqnarray}\label{final}
{\cal A}_V&=&-\frac{\alpha\Gamma(4\gamma+1)}
{\pi\gamma\Gamma^2(2\gamma+1)}
\int_0^\infty\frac{dt}{t}\sum_{\lambda=\pm
1}\int_{0}^{\infty}\!\!\!\int_{0}^{\infty}\frac{ds_1ds_2\, \exp[\,
2iZ\alpha\lambda (s_1+s_2)]}{(\sinh s_1\sinh
s_2)^{2\gamma}} \nonumber\\
&&\times\Biggl\{\left[ \frac{(1-\coth s_1\coth
s_2)}{D_-^{4\gamma}}-
\frac{(1+\coth s_1\coth s_2)}{D_+^{4\gamma}}\right]\nonumber\\
&&+\frac{1}{4\gamma-1}\left[ \frac{|t-1/t|(Q+\coth s_1\coth
s_2)}{D_-^{4\gamma-1}}-
\frac{(t+1/t)(-Q+\coth s_1\coth s_2)}{D_+^{4\gamma-1}}\right]+\nonumber\\
&&+\frac{1}{(4\gamma-1)(4\gamma-2)}\left[
\frac{Q(3-t^2-1/t^2)+\coth s_1\coth s_2}{D_-^{4\gamma-2}}-
\frac{Q(-3-t^2-1/t^2)+\coth s_1\coth
s_2}{D_+^{4\gamma-2}}\right]-\nonumber\\
&&-\frac{3Q}{(4\gamma-1)(4\gamma-2)(4\gamma-3)}\left[
\frac{|t-1/t|}{D_-^{4\gamma-3}}- \frac{(t+1/t)}{D_+^{4\gamma-3}}\right]-\nonumber\\
&&-\frac{3Q}{(4\gamma-1)(4\gamma-2)(4\gamma-3)(4\gamma-4)}\left[
\frac{1}{D_-^{4\gamma-4}}-
\frac{1}{D_+^{4\gamma-4}}\right]\Biggr\}\, .
\end{eqnarray}
The following notation  are used:
\begin{eqnarray}\label{bD}
Q&=&(1-\gamma)\left[(1+\gamma)\coth s_1\coth s_2-(1-\gamma) -
iZ\alpha\lambda (\coth s_1+\coth s_2)\right]
 ,\nonumber\\
D_-&=&t\coth s_1+(1/t)\coth s_2+|t-1/t|\, ,\, D_+=t\coth
s_1+(1/t)\coth s_2+(t+1/t)\, .
\end{eqnarray}

\begin{figure}[h]
\centering
\includegraphics[height=180pt,keepaspectratio=true]{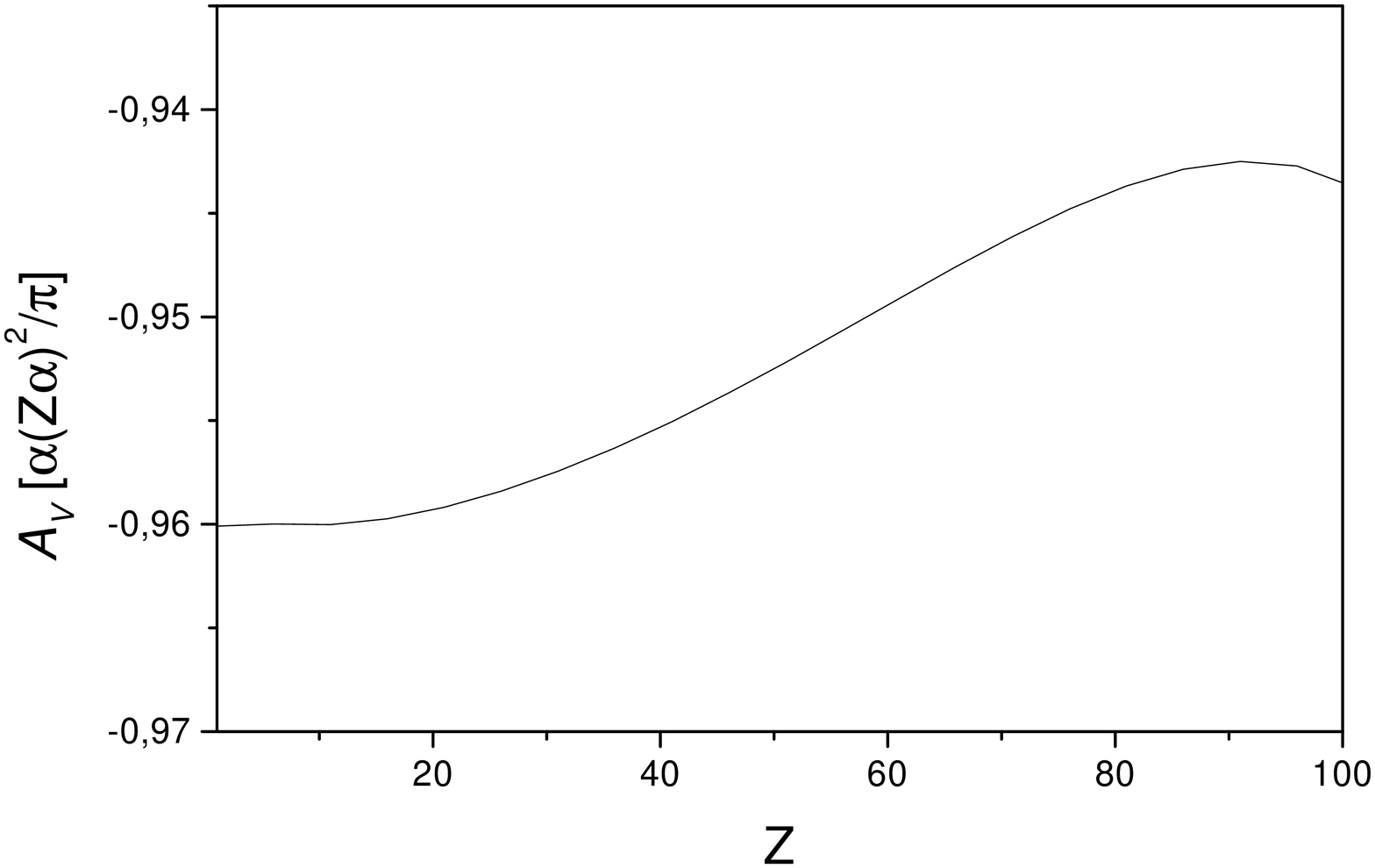}
\caption{\it The function ${\cal A}_{V}$ calculated in all orders in
$Z\alpha$. Value of the function is given in units
$\frac{\alpha(Z\alpha)^2}{\pi}$.
} \label{Fig4}
\end{figure}

The part of the vertex that is independent of $Z\alpha$ requires
the ultraviolet regularization and hence is dependent on the
regularization parameter. In fact we have put the regularization
parameter to be equal to $1/r_0$. The corresponding
$Z$-independent term in Eq.(\ref{final}) equals to $\alpha/2\pi$.
One should also take into account the term that provides the
correct normalization of the total wave function, see e.g.
Ref.\cite{Pach96},
\begin{eqnarray}\label{ren}
\delta_N&=&-\alpha\int\limits_{-\infty}^{\infty}\!\!
d\epsilon\int\!\!\!\int\!\!\!\int\!\!\! d\bm r_1d\bm r_2d\bm
r_3\,\bar\psi_p({\bm r_2})\gamma_\mu {\cal G}(\bm
r_2,\bm r_3|\,i\epsilon)\nonumber\\
&&\times \gamma^0{\cal G}(\bm r_3,\bm
r_1|\,i\epsilon)\gamma^\mu\psi_p(\bm r_1)D(\bm r_2,\bm
r_1|i\epsilon)\nonumber\\
&& + (\psi_p\to\psi_s)\,\, .
\end{eqnarray}
Straightforward calculation of this contribution with logarithmic
accuracy gives $-\alpha\,L/2\pi$. Note that there are no
$Z$-dependent logarithmic terms in (\ref{ren}). As a result the
"normalization" contribution cancels out exactly the
$Z$-independent term in ${\cal A}_V$. In essence this cancellation
is a direct consequence of the Ward identity. The leading term
of $Z\alpha$-dependent part of  ${\cal A}_V$ at $Z\alpha\ll 1$
reads
\begin{eqnarray}\label{fanswer}
{\cal A}_V
=-\frac{\alpha(Z\alpha)^2}{\pi}\left(\frac{17}{4}-\frac{\pi^2}{3}\right)\quad.
\end{eqnarray}
Numerical calculation of integrals in Eq. (\ref{final}) together with
account of the normalization contribution  (\ref{ren}) gives the function
${\cal A}_{V}$ exactly in $Z\alpha$. This function is plotted in
Fig. \ref{Fig4}

\begin{figure}[h]
\centering
\includegraphics[height=180pt,keepaspectratio=true]{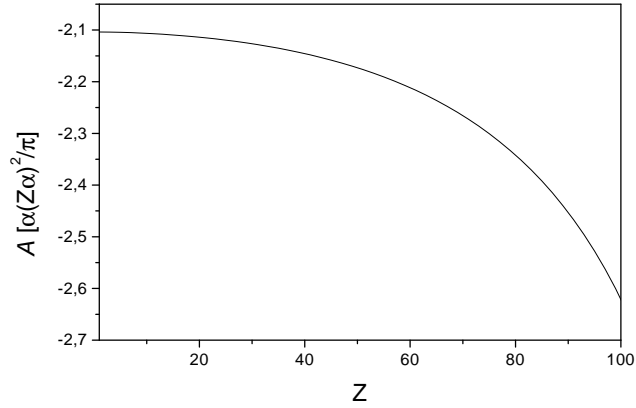}
\caption{\it The function ${\cal A}$ calculated in all orders in
$Z\alpha$. Value of the function is given in units
$\frac{\alpha(Z\alpha)^2}{\pi}$.
} \label{Fig5}
\end{figure}

Equations  (\ref{munu}) and (\ref{fanswer}) give the final result for
${\cal A}$ at $Z\alpha\ll 1$
\begin{eqnarray}\label{finallog}
{\cal A}= {\cal A}_{SE}+{\cal A}_V=
-\frac{\alpha(Z\alpha)^2}{\pi}\left(\frac{15}{4}-\frac{\pi^2}{6}\right)\quad.
\end{eqnarray}
The function ${\cal A}(Z\alpha)$ with account of all orders in $Z\alpha$
is plotted in Fig. \ref{Fig5}

\section{Linear in $Z\alpha$ radiative corrections}\label{PNClinear}

\begin{figure}[h]
\centering
\includegraphics[height=55pt,keepaspectratio=true]{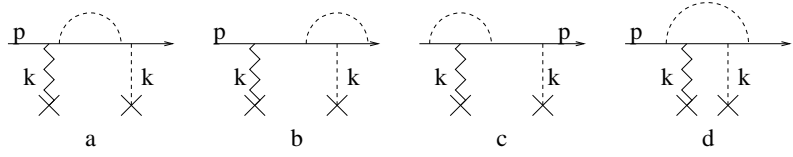} \caption{\it
{Self-energy and vertex radiative corrections to the forward scattering
amplitude. The zigzag line
denotes Z-boson and the dashed line denotes photon. The cross
denotes the nucleus. Contribution of each diagram must be doubled
because of permutation of Z-boson and Coulomb lines.}}
 \label{Fig6}
\end{figure}

In the leading $\alpha(Z\alpha)$ approximation, it is convenient
to derive the non-logarithmic term of the radiative correction
using the  effective operator approach, see e. g. Ref.
\cite{Eides}. In this approach the corrections under discussion
coincide with those for the ${\bm \sigma}\cdot {\bm p}$ structure
in the amplitude of low-energy forward scattering. Diagrams for
the scattering amplitude are shown in Fig.\ref{Fig6}. Below we
double the contribution of each diagram  because of permutation of
Z-boson and Coulomb lines.
In these amplitudes the momentum $|\bm p|$ of the external
electron is small compared to the momentum of the Coulomb quantum,
$k\sim m$. Renormalization procedures for the self-energy operator
and for the photon vertex operator are standard ones. For the
Z-boson vertex we have to set the vertex correction equal to zero
at $k=0$ because as a reference point we use result \cite{Mar1}
that is normalized at $k=0$.
 To minimize the infrared problems we perform calculations in the
Fried-Yennie gauge \cite{FY} where the photon propagator is of the
form
\begin{eqnarray}\label{Yennie}
D_{\mu\nu}(q)=\frac{g_{\mu\nu}q^2+2q_\mu q_\nu}{(q^2+i0)^2} \quad.
\end{eqnarray}
This gauge does not require an introduction of the photon mass for
infrared regularization. In this gauge the renormalized
self-energy operator has the form \cite{Eides}
\begin{equation}\label{SER}
\bm \Sigma(p)=-\frac{3\alpha{\hat p}\,({\hat p}-m)^2}{4\pi
m^2(1-\rho)}\left(1+\frac{\rho}{1-\rho}\ln\rho\right),
\end{equation}
where $\rho=1-p^2/m^2$. Calculating  the diagram
Fig.\ref{Fig6}(a) with this self-energy operator we obtain the
following  contribution to the function ${\cal B}$ defined in Eq.
(\ref{AB})
\begin{equation}\label{Ba}
{\cal B}_a=\frac{5}{4}\alpha(Z\alpha)\, .
\end{equation}
A straightforward calculation leads to the following expression
for the renormalized correction to the  photon vertex operator

\begin{eqnarray}
\label{LaR}
&&\Lambda_{\alpha}
=\frac{\alpha}{4\pi}\left\{-3\gamma_{\alpha} \int_0^1dy\left(\ln
[1+\tau y(1-y)]+\frac{1}{2}\right)
+\int_0^1dy\int_0^1\frac{dx}{m^2T}\right.\\
&& \times\left[x^2\left(b^2\gamma_{\alpha}+2{\hat
b}\gamma_{\alpha}{\hat b}\right) +2x\left(\gamma_{\alpha}{\hat
p_2}{\hat b}+ {\hat b}{\hat p_1}\gamma_{\alpha}- {\hat
b}\gamma_{\alpha}{\hat p_1}- {\hat p_2}\gamma_{\alpha}{\hat b}
+6mb_{\alpha}\right) +(x-2)\gamma_{\mu}({\hat
p_1}+m)\gamma_{\alpha}({\hat p_2}+m)\gamma_{\mu}
\right]\nonumber\\
&&+\left.\int_0^1dy\int_0^1\frac{(1-x)xdx}{m^4T^2} \left[2{\hat
b}({\hat p_1}+m)\gamma_{\alpha}({\hat p_2}+m){\hat b}\right]
\right\}.\nonumber
\end{eqnarray}
Here the incoming electron has the momentum $p_2$ and outgoing
electron the momentum $p_1=p_2+k$,
 $$b=[p_1(1-y)+p_2y]\, ,\quad T=x+\tau y(1-xy)+\lambda
 y(1-x),\quad
 \tau=\frac{{\bm k}^2}{m^2},\quad
 \lambda=\frac{2{\bm p}\cdot{\bm k}}{m^2}\,\,.$$
Performing calculations with this vertex function we find the
contribution of the diagram Fig.\ref{Fig6}(b)
\begin{equation}\label{Bb}
{\cal B}_b=\alpha(Z\alpha)\left(-\frac{7}{12}-\frac{8}{3}\ln
2\right)\, .
\end{equation}
The renormilized correction to the $Z$-boson vertex reads
\begin{eqnarray}
\label{LaR5}
&&\L_{\alpha}=\frac{\alpha}{4\pi}\left\{-\gamma_{\alpha}
\int_0^1dy\left(3\ln [1+\tau y(1-y)]-\frac{1}{2}\right)
+\int_0^1dy\int_0^1\frac{dx}{m^2T}\right.\\
&& \times\left[x^2\left(b^2\gamma_{\alpha}+2{\hat
b}\gamma_{\alpha}{\hat b}\right) +2x\left(\gamma_{\alpha}{\hat
p_2}{\hat b}+ {\hat b}{\hat p_1}\gamma_{\alpha}- {\hat
b}\gamma_{\alpha}{\hat p_1}- {\hat p_2}\gamma_{\alpha}{\hat b}
+m({\hat b}\gamma_{\alpha}-\gamma_{\alpha}{\hat b})\right)
\right.\nonumber\\
&&\left.+(-2+x)\gamma_{\mu}({\hat p_1}+m)\gamma_{\alpha}({\hat
p_2}-m)\gamma_{\mu}\right]+\left.\int_0^1dy\int_0^1\frac{(1-x)xdx}{m^4T^2}
\left[2{\hat b}({\hat p_1}+m)\gamma_{\alpha}({\hat p_2}-m){\hat
b}\right] \right\}\gamma_5.\nonumber
\end{eqnarray}
Calculation with this vertex function gives the following
contribution of the diagram Fig.\ref{Fig6}(c)
\begin{equation}\label{Bc}
{\cal B}_c=\alpha(Z\alpha)\left(\frac{1}{4}-\frac{10}{3}\ln
2\right)\, .
\end{equation}
The last diagram, Fig.\ref{Fig6}(d), does not require any
renormalization. A straightforward calculation gives
\begin{equation}\label{Bd}
{\cal B}_d=\alpha(Z\alpha)\left(-\frac{3}{2}+4\ln 2\right)\, .
\end{equation}
Altogether, the result  for the non-logarithmic term in
Eq.(\ref{AB}) reads
\begin{equation}\label{nlfinal}
{\cal B}={\cal B}_a+{\cal B}_b+{\cal B}_c+{\cal B}_d=
-\alpha(Z\alpha)\left(\frac{7}{12}+2\ln 2\right).
\end{equation}

Thus, according to Eqs. (\ref{AB}), (\ref{finallog}), and
(\ref{nlfinal}) total relative correction in the leading
$Z\alpha$-approximation reads
\begin{eqnarray}\label{finalPNC}
\delta_{ef}&=&-\alpha\left[(Z\alpha)\left(\frac{7}{12}+2\ln2\right)
+\frac{(Z\alpha)^2}{\pi}\left(\frac{15}{4}-\frac{\pi^2}{6}\right)
\ln(b\lambda_C/r_0) \right]\quad.
\end{eqnarray}
The curve corresponding to this formula is shown in Fig.\ref{Fig7}
by the dotted line pnc1. By the dashed line pnc2 we show the
correction $\delta_{ef}$ calculated with the exact in $Z\alpha$
function ${\cal A}$.
The solid line pnc in the same
Fig.\ref{Fig7} shows the total radiative correction to the PNC
effect that includes both $\delta_{ef}$ and $\delta_{b}$
(Eq.(\ref{db}), Fig.\ref{Fig1}(b)).
The leading unaccounted contribution in $\delta_{ef}$ is of the
order of $\sim Z^2\alpha^3/\pi$. For Cs (Z=55) this gives about 5\%
uncertainty in $\delta_{ef}$.

\section{Radiative corrections to the finite-nuclear-size effect}\label{FNS}

\begin{figure}[h]
\centering
\includegraphics[height=180pt,keepaspectratio=true]{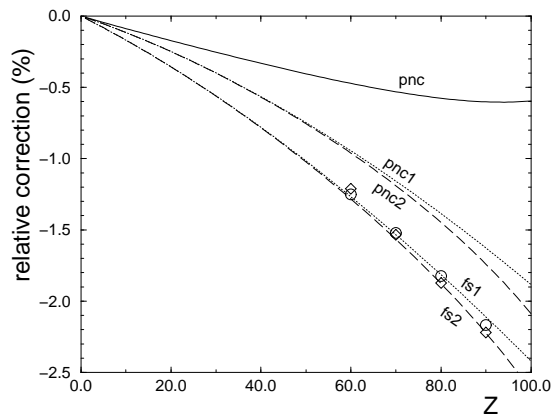}
\vspace{-10pt} \hspace{0pt} \caption{\it { Relative radiative
corrections (\%) for the PNC and for the finite-nuclear-size
effects versus the nuclear charge $Z$. The dotted line pnc1
shows the correction $\delta_{ef}$ (Fig.\ref{Fig1}(e,f))
when both logarithmic and nonlogarithmic terms are calculated in leading
in $Z\alpha$ orders, see Eq.(\ref{finalPNC}).
The same correction, but with logarithmic term calculated {\it exactly}
in $Z\alpha$, is shown by the dashed line pnc2.
The solid line pnc shows the total radiative
correction to the PNC effect that includes both $\delta_{ef}$ and
$\delta_{b}$ (Eq.(\ref{db}), Fig.\ref{Fig1}(b)).
The dotted line fs1 shows the finite nuclear size correction $\Delta_s$
when both logarithmic and nonlogarithmic terms are calculated in leading
in $Z\alpha$ orders, see Eq.(\ref{finaFNSS}).
The same correction, but with logarithmic term calculated {\it exactly}
in $Z\alpha$, is shown by the dashed line fs2.
Results of computations of $\Delta_s$ for $1s$ and $2s$ states
\cite{CJS93,JS} are shown by circles and diamonds, respectively.
 }}
\label{Fig7}
\end{figure}

The radiative shift of the atomic energy levels (Lamb shift)
depends on the finite nuclear size. This correction has a very
similar structure to that of the PNC radiative correction since
the effective sizes of the perturbation sources in both cases are
much smaller than $\lambda_C$. The self-energy and the vertex
corrections to the finite-nuclear-size effect (SEVFNS) for
$s_{1/2}$-state have been calculated earlier analytically in order
$\alpha(Z\alpha)$ in Refs.\cite{Pach93,PG}. The corrections for
$1s_{1/2}$-, $2s_{1/2}$-, and $2p_{1/2}$-states have been
calculated numerically exactly in $Z\alpha$ in
Refs.\cite{CJS93,Blun92,LPSY}. However, the structure of higher in
$Z\alpha$ corrections and their logarithmic dependence on the
nuclear size has not been understood.
 We have applied our approach to the SEVFNS problem
 and found the following expression for the $s_{1/2}$-state
 relative correction in the leading $Z\alpha$-approximation
\begin{eqnarray}\label{finaFNSS}
\Delta_s=-\alpha\left[(Z\alpha)\left(\frac{23}{4}-4\ln2\right)+
\frac{(Z\alpha)^2}{\pi}\left(\frac{15}{4}-\frac{\pi^2}{6}\right)
\ln(b\lambda_C/r_0) \right]\quad.
\end{eqnarray}
Linear in $Z\alpha$ term agrees with results of Refs.
\cite{Pach93,PG}. The correction $\Delta_s$ given by Eq.
(\ref{finaFNSS}) is shown in Fig. \ref{Fig7}
by the dotted line fs1. By the dashed line fs2 we show the
correction  $\Delta_s$ calculated
with the exact in $Z\alpha$ function ${\cal A}$.
Results of the computations \cite{CJS93,JS} for $1s$ and $2s$
states are shown by circles and diamonds, respectively. The
agreement is excellent.

Note that logarithmic terms ($\ln(b\lambda_C/r_0)$) in $\delta_{ef}$ and $\Delta_s$
coincide. This statement is valid in all orders in $Z\alpha$.
Moreover, the logarithmic term in the SEVFNS correction $\Delta_p$
for $p_{1/2}$ state is also equal to that in $\delta_{ef}$ and
$\Delta_s$. The reason for this equality is very simple. The
logarithmic terms come from small distances ($r << \lambda_C$)
where the electron mass can be neglected. When the mass is
neglected the relative matrix elements for the PNC radiative
correction and for SEVFNS are equal.

\noindent
The correction $\Delta_p$ is calculated in our work \cite{com}.
The result reads
\begin{eqnarray}
\label{Dp}
\Delta_p=-\alpha\left[-\frac{8}{9\pi}
\left(\ln\frac{1}{(Z\alpha)^2}+0.910\right)
+ 2.75(Z\alpha)+\frac{(Z\alpha)^2}{\pi}\left(\frac{15}{4}-\frac{\pi^2}{6}\right)
\ln(b\lambda_C/r_0) \right]\, .
\end{eqnarray}
Structure of this correction is qualitatively different from that
of $\delta_{ef}$ and $\Delta_s$, the expansion starts from
$\alpha\ln(1/Z\alpha)$ term, while in  $\delta_{ef}$ and
$\Delta_s$ it starts from $Z\alpha^2$.  This difference is due to
the different infrared dynamics, see discussion in \cite{com}.

It has been  suggested in Ref. \cite{K} that the relation
$\delta_{ef}=(\Delta_s+\Delta_p)/2$ is valid. 
 Our results (\ref{finalPNC}), (\ref{finaFNSS}), and (\ref{Dp}) clearly
disagree with this relation.
Although, due to accidental compensations, the relation is more or less valid 
numerically around $Z\approx 57$, we strongly insist that there is no valid 
justification for the relation.
The "derivation" in Ref. \cite{K} is based on an assumption
that there is a gauge in which the vertex contributions to $\delta_{ef}$, $\Delta_s$, 
and $\Delta_p$ vanish simultaneously. This assumption is obviously wrong, and
this is why the above relation is wrong.

\section{Conclusion}\label{Conclusion}
Now we can perform a consistent analysis of the experimental data
on atomic parity violation since all the contributions are known.
In our analysis for Cs we include the theoretical value of the PNC
amplitude from Refs. \cite{Dzuba2,Blundell,Kozlov,DFG}
\begin{equation}
\label{EP} E_{PNC}= 0.908 (1\pm 0.005) \ 10^{-11} \ i e  a_B
(-Q_W/N),
\end{equation}
as well as the $-0.61\%$ correction due to the Breit interaction
\cite{DHJS}, the $-0.85\%$ radiative correction calculated in the
present work, the $+0.42\%$ vacuum polarization correction
\cite{W,MiSu}, the $-0.2\%$ neutron skin correction \cite{Der1},
the $-0.08\%$ correction due to the renormalization of $Q_W$ from the
atomic momentum transfer $q\sim 30$MeV down to $q=0$ \cite{MiSu},
and the $+0.04\%$ contribution from the electron-electron weak
interaction \cite{MiSu}.
The theoretical uncertainty from the $Z^2\alpha^3/\pi$ term unaccounted
in the present calculation is about $0.05 - 0.1\%$.
 In Ref. \cite{Cs} the ratio
$E_{PNC}/\beta$ has been measured with 0.35\% accuracy. Here
$\beta$ is the vector transition polarizability. An analysis of the
recent data on $\beta$ has been performed in Ref.\cite{DFG}. We
use the value $\beta= 26.99(5)a_B^3$ obtained in \cite{DFG}.
Combining all these results we obtain the following value of the
nuclear weak charge $Q_W$ at zero momentum transfer
\begin{equation}
\label{QCs} Cs: \quad Q_W=-72.81 \pm (0.28)_{ex}\pm(0.36)_{theor}.
\end{equation}
This value agrees with prediction of the standard model,
$Q_W=-73.09\pm 0.03$, see Ref. \cite{RPP}. We have used the
neutron skin correction  in our analysis. However, in our opinion,
status of this correction is not quite clear because  data on
the neutron distribution used in Ref. \cite{Der1} are not quite
consistent with  data on the neutron distributions obtained from
proton scattering, see e. g. Ref. \cite{LIYAF}.

In the analysis for Tl we have included the theoretical value of
the PNC amplitude from Refs. \cite{Dzuba1}, as well as the $-0.88\%$
correction due to the Breit interaction \cite{rescale}, the $-1.48$\%
radiative correction calculated in the present work, the $+0.90\%$
vacuum polarization correction \cite{MiSu}, the $-0.2\%$
neutron skin correction, the $-0.08\%$ correction due to the
renormalization of $Q_W$ from the atomic momentum transfer $q\sim
30$MeV down to $q=0$ \cite{MiSu}, and the $+0.01\%$ contribution from
the electron-electron weak interaction \cite{MiSu}. Using these
theoretical results we obtain from the data \cite {Tl} the
following value of the nuclear weak charge $Q_W$ at zero momentum
transfer
\begin{equation}
\label{QTl} Tl: \quad Q_W=-116.8 \pm (1.2)_{ex}\pm(3.4)_{theor}.
\end{equation}
This  agrees with prediction of the standard model, $Q_W=-116.7\pm
0.1$, see Ref. \cite{RPP}. Values of the weak charge $Q_W$ obtained
in the present work differ a little from that reported in \cite{Mil}.
There are two reasons for the difference, a) in the present work we
have used a slightly different value for Cs vector transition
polarizability, this value is probably more accurate,
b) in the present work calculation of the logarithmic
term in the radiative correction is performed exactly in $Z\alpha$,
while in \cite{Mil} we have used only the leading term of $Z\alpha$-expansion.

Concluding, we have calculated the radiative corrections to the effect of
atomic parity violation, the corrections are enhanced by the collective
electric field of the nucleus . This calculation has allowed us to
perform a consistent analysis of the experimental data on the nuclear weak
charge. Agreement with the standard model is within $0.6\sigma$.

We would like to thank V. A. Dzuba for helpful discussions. A.I.M
gratefully acknowledge the School of Physics at the University of
New South Wales for  warm hospitality and financial support during
a visit.


\begin{thebibliography}{99}
\bibitem{rescale} We found this value by rescaling the Breit
correction for Cs \cite{DHJS}, see discussion in Refs.
\cite{Sushkov,MiSu}.
\end{thebibliography}
\end{document}